\documentstyle[preprint,aps,epsf]{revtex}
\newif\iftightenlines\tightenlinesfalse
\tightenlines\tightenlinestrue

\def\eslt{\not\!\!{E_T}}
\def\to{\rightarrow}

\def\tu{\tilde u}

\def\tst{\tilde t}
\def\ttau{\tilde \tau}

\def\tg{\tilde g}

\def\tell{\tilde\ell}
\def\tq{\tilde q}
\def\tw{\widetilde W}
\def\tz{\widetilde Z}
\begin{document}
\draft
\preprint{\vbox{\baselineskip=14pt%
   \rightline{FSU-HEP-000708}
   \rightline{UH-511-965-00}
}}
\title{REACH OF THE CERN LHC FOR THE MINIMAL \\
ANOMALY-MEDIATED SUSY BREAKING MODEL}
\author{Howard Baer$^1$, J.~K.~Mizukoshi$^2$ and Xerxes Tata$^2$}
\address{
$^1$Department of Physics,
Florida State University,
Tallahassee, FL 32306 USA
}
\address{
$^2$Department of Physics and Astronomy,
University of Hawaii,
Honolulu, HI 96822, USA
}
\date{\today}
\maketitle
\begin{abstract}
We examine the reach of the CERN LHC $pp$ collider for supersymmetric
models where the dominant contribution to soft SUSY breaking parameters
arises from the superconformal anomaly. In the simplest viable anomaly
mediated SUSY breaking (AMSB) model, tachyonic slepton squared masses
are made positive by adding a universal contribution $m_0^2$ to all
scalars.  We use the event generator ISAJET to generate AMSB signal
events as a function of model parameter space. Assuming an integrated
luminosity of 10 fb$^{-1}$, the LHC can reach to values of $m_{\tg}\sim
2.3$ TeV for low values of $m_0$, where the dilepton plus jets plus
$\eslt$ channel offers the best reach. For large $m_0$, the best
signature is typically 0 or 1 isolated lepton plus jets plus $\eslt$; in
this case the reach is typically diminished to values of $m_{\tg}\sim
1.3$ TeV.
The presence of terminating tracks in a subset of signal
events could serve to verify the presence of a long lived lightest chargino
which is generic in the minimal AMSB model.
\end{abstract}

\medskip

\pacs{PACS numbers: 14.80.Ly, 13.85.Qk, 11.30.Pb}


The hypothesis of weak scale supersymmetry (SUSY) resolves several 
problems in particle physics, most notable of which is the 
gauge hierarchy problem. A consequence of the hypothesis is that 
supersymmetric matter should exist with masses typically 
less than $\sim 1$ TeV\cite{review}.
These SUSY particles are expected to be produced with large cross sections
at the CERN Large Hadron Collider (LHC). Specific SUSY particle
signatures, however, depend on assumptions underlying the 
mechanism of SUSY breaking  and its mediation to the superpartners of
ordinary particles.

Since dynamical SUSY breaking at the TeV scale does not appear to be
possible within the framework of the Minimal Supersymmetric Standard
Model (MSSM), a ``hidden sector'' where SUSY breaking occurs is
generally assumed.  SUSY breaking is then communicated to the visible
sector via messenger interactions. In supergravity models, gravity acts
as the messenger, and TeV level sparticle masses can arise if hidden
sector SUSY breaking vevs of order $M\sim 10^{11}$ GeV can be arranged.
Generically in these models, the induced soft SUSY breaking (SSB) parameters
lead to problems with FCNCs and CP violating processes. 

Recently, it has been noted that in supergravity models, there exist
loop-suppressed contributions to soft SUSY breaking masses arising from
the superconformal anomaly\cite{rs,glmr}. Usually, these contributions
are sub-dominant compared to tree level supergravity induced mass terms.
However, in the higher dimensional universe scenario of Randall and
Sundrum~\cite{rs}, 
SUSY breaking can take place in a brane separated from the visible sector
brane by a large distance (the sequestered sector model). 
In this case, tree level supergravity masses do not occur, and
the anomaly mediated SUSY breaking (AMSB) mass contributions dominate.

The AMSB soft SUSY breaking gaugino masses have the special property
that they are proportional to the beta functions of their corresponding 
gauge group:
\begin{equation}
M_i=\frac{\beta_i}{g_i} m_{3/2} ,
\end{equation}
where $i=3,2,1$ for gauge group $SU(3)$, $SU(2)$ and $U(1)$, $g_i$ is
the corresponding gauge coupling, and $m_{3/2}$ is the gravitino mass.
Similar formulae for SSB scalar squared masses and/or $A$ parameters
were worked out in \cite{rs,glmr,wells,fm}; the formulae we use are
presented in \cite{bdqt} in the notation of the event generator program
ISAJET~\cite{isajet} that we use for our simulation, and will not be
repeated here.  The characteristic features of the SSB mass terms are
the following.
\begin{itemize}
\item Since scalar masses are proportional to their beta functions, they
only depend on
their respective gauge and Yukawa couplings. This guarantees 
universality between
first and second generation scalar flavors, thus providing a solution to
the SUSY flavor problem. 
It has been argued~\cite{rs} that these models also have the potential for  
resolving the SUSY $CP$ problem.

\item The gaugino masses occur in the approximate ratio $M_1:M_2:M_3 =
2.8:1:7.1$, so that the $SU(2)$ gaugino is the lightest.  This fact has
profound consequences for phenomenology. The lightest neutralino $\tz_1$
is dominantly a wino, and there is only a tiny mass gap between
$m_{\tw_1}$ and $m_{\tz_1}$. Including loop corrections,
$m_{\tw_1}-m_{\tz_1}\agt 160$ MeV, and the decay $\tw_1^\pm\to \pi^\pm
\tz_1$ dominates, although the pion energy will be very soft.  This
implies that the $\tw_1$ will give rise to nearly invisible decay
products.  The lightest chargino $\tw_1$ is long lived, with values of
$c\tau$ typically a few $cm$.  It is thus possible that the $\tw_1$
leaves an observable but terminating track, but whether or not this is
detectable depends crucially on the detector apparatus.

\item Since the Standard Model (SM) $U(1)_Y$ and $SU(2)_L$ gauge groups are 
not asymptotically
free, negative squared masses for sleptons are predicted, leading to
tachyonic states. This is viewed as the most severe shortcoming of the
AMSB model, and numerous proposals have been put forth to
ameliorate the situation\cite{other}. 
Usually, these involve introducing new fields
at intermediate scales. A simple proposal, dubbed the minimal AMSB
(mAMSB) model, which is phenomenologically
acceptable, is that all scalars receive an additional universal contribution 
$m_0^2$\cite{rs,wells,fm}, which of course, preserves the equality
of masses for scalars with the same gauge quantum numbers.
\end{itemize}

Previous studies have addressed discovery of AMSB at collider
experiments\cite{fmrss,wells,fm,frank,probir}. In \cite{fmrss}, several
new triggers for the Fermilab Tevatron $p\bar{p}$ collider experiments
were suggested that exploit the highly ionizing track expected from the
chargino. In \cite{wells}, it is suggested that Tevatron experiments
look for isolated dilepton plus missing energy events which contain in
addition displaced vertices from chargino decays. In \cite{fm}, it is
emphasized that naturalness constraints imply light chargino masses
which should generally be accessible to Tevatron searches.  In
Ref. \cite{frank}, the AMSB scenario was examined with regard to LHC
searches. Here, a detailed analysis of one case study was performed, and
it was shown that many of the techniques of reconstructing sparticle
masses for mSUGRA models\cite{ph} could be applied to the AMSB
case. Finally, Ref.~\cite{probir} discusses slepton signals at a high
energy $e^+e^-$ colliders within this framework.

An important issue in SUSY collider phenomenology is whether SUSY can be
discovered or excluded over the entire range of reasonable model
parameters. Although SUSY may be discovered at the Tevatron, for all
distinct scenarios of SUSY breaking, there exist reasonable ranges of
model parameter choices where no sparticles will be
discovered\cite{run2}.  However, the CERN LHC collider is frequently
regarded as the definitive machine where weak scale SUSY will either be
discovered or essentially disproved. The reach of the CERN LHC has been
examined for the mSUGRA model for both high and low values of
$\tan\beta$, and with and without $R$-parity violation\cite{lhc,cms}.
The LHC reach for models with gauge-mediated SUSY breaking (GMSB) has been
explored for various model lines giving rise to distinct collider
signatures, and again the reach has been found to be
considerable\cite{gmsb}.  The unique sparticle mass spectrum and gaugino
content of the LSP in AMSB models necessitates a distinct analysis for
this class of models.  In this paper, we present results showing the
reach of the CERN LHC for the mAMSB model.

The mAMSB model has been presented in Ref.~\cite{wells,fm,bdqt}.
Within this framework, all
sparticle masses and couplings can be calculated in
terms of the parameter set:
\begin{equation}
m_0,\ m_{3/2},\ \tan\beta\ {\rm and}\ sign(\mu ) .
\end{equation}
The mAMSB model has recently~\cite{bdqt} been embedded
in the event generator ISAJET. In ISAJET, the weak scale 
gauge and Yukawa couplings are used as inputs to determine the approximate
GUT scale. At $M_{GUT}$, the mAMSB SSB masses are calculated, and used as 
boundary conditions to determine their weak scale counterparts. An iterative
solution is employed in ISAJET 7.51, including 2-loop RGEs for
couplings and SSB masses, and minimization of the renormalization group 
improved 1-loop 
effective potential at an optimized scale choice which effectively includes
dominant 2-loop terms. Electroweak symmetry is broken radiatively (REWSB).

As far as LHC signatures go, the feature that most distinguishes the
mAMSB from other frameworks (mSUGRA, GMSB) where the gaugino masses
unify at $M_{GUT}$ is that the $SU(2)$ gaugino is much lighter than the
bino. Since $\mu$ is typically large, the lighter chargino ($\tw_1$) and
neutralino ($\tz_1$) are essentially wino-like with $m_{\tz_1}$ only
slightly lighter than $m_{\tw_1}$, while the second lightest neutralino
($\tz_2$) is essentially a bino. This has a large impact on the cascade
decay patterns of gluinos and squarks. In particular, since gluinos and
$\tq_L$ dominantly decay to $SU(2)$-like gauginos~\cite{bbkt}, within
the mAMSB framework their cascade decays do not result in hard
leptons. In contrast, $\tq_R$ now mainly decays to the hypercharge
gaugino, {\it i.e.} via $\tq_R \to q\tz_2$, so that $\tq_R$ production
is a potential origin of leptonic signals. This is in sharp contrast to
the mSUGRA or GMSB frameworks, where these signals mainly come from
$\tg$ and $\tq_L$ production. The jet-free trilepton signature from
$\tw_1\tz_2$ production is also expected to be small, partly because the
$\tw_1\tz_2$ cross section is small, but mainly because the $\tw_1$ does
not decay leptonically. Thus if squarks are heavy, multilepton signals
are expected to be relatively suppressed within the mAMSB framework.

We now turn to the reach of the LHC. We use the generic analysis detailed in
our studies~\cite{lhc} of the reach within the mSUGRA framework to
obtain the reach for the present case.
For each point in a grid in the mAMSB 
parameter space, ISAJET calculates the relevant sparticle mass
spectrum, and all sparticle and Higgs boson branching ratios. ISAJET then 
generates all SUSY $2\to 2$ production processes, followed by cascade decays,
QCD radiation in the parton shower approximation, hadronization and
beam remnant evolution. The final state particles are passes to a toy 
detector simulation represenative of LHC detectors. Calorimeter
extent, cell size and energy smearing and jet finding algorithm 
are as given in Ref. \cite{lhc},
and will not be repeated here. 

In each signal event, we require 
the presence of {\it i}) at least two jets, each with $E_T(jet)>E_T^c$,
where $E_T^c$ is a floating cut value designed to give some optimization of 
cuts throughout parameter space. We use values of $E_T^c=100-500$ GeV,
in steps of 100 GeV. We further require {\it ii}) $\eslt >E_T^c$ and 
{\it iii}) $S_T>0.2$, where $S_T$ is the transverse sphericity. For 
multilepton events, we require each lepton $p_T(\ell )>10$ GeV, $|\eta_\ell |
<2.5$ and visible activity in a cone of size $\Delta R=0.3$ about the lepton
direction to be less than $E_T(cone)=5$ GeV (isolation). 
In addition, for events with 
two or more isolated leptons, the two hardest leptons in $p_T$ should have
$p_T(\ell )>20$ GeV. In the special case of single isolated lepton events,
we require $p_T(\ell )>20$ GeV, and transverse mass $m_T(\ell ,\eslt )>100$ 
GeV to suppress backgrounds from $W$ production. The signal events are
then classified according to the number of isolated leptons: $0\ell$
for multijet plus missing energy events with no isolated leptons, 
$1\ell$ for single isolated 
lepton events, $OS$ for events containg two opposite sign isolated leptons,
$SS$ for events with two same-sign isolated leptons and $3\ell$ for
events containing three isolated leptons.

SM background rates arising from $t\bar t$ production, QCD multijet
production, $W$ or $Z$ plus jets production and vector boson pair
production have been calculated in Ref.~\cite{lhc} as a function of
$E_T^c$ for each of the signal channels. We use the background from this
paper~\footnote{We caution the reader that these were obtained using
CTEQ2L structure functions while for the signal we are using CTEQ3L
structure functions.} further requiring that $E_T^c \le 400$~GeV
(300~GeV) for the OS and SS (3$\ell$) channels.  A signal is regarded as
visible in a particular channel if, for any value of $E_T^c$, ({\it
i})~it is $5\sigma$ above the SM background for an integrated luminosity
of 10 fb$^{-1}$, ({\it ii})~the signal cross section is larger than 20\%
of the corresponding background cross section, and ({\it iii}) there
should be at least 5 signal events for an integrated luminosity of
10~fb$^{-1}$.

Our first results are shown in Fig.~\ref{amsb1}, where we show the
$m_0\ vs.\ m_{3/2}$ parameter plane for $\tan\beta =3$ and $\mu <0$.
The upper-left  shaded region is excluded because the $\ttau_1$ slepton
becomes the LSP. The low $m_{3/2}$ shaded region is excluded by
LEP2 searches for charginos in the mAMSB scenario: $m_{\tw_1}> 86$ GeV
assuming a heavy sneutrino mass\cite{grenier}~\footnote{If the sneutrino
is light the negative interference caused by the $t$-channel sneutrino
exchange amplitude diminishes the limit 
to 68 GeV. The region of parameters where this might be relevant is
excluded by the $h$ mass limit.}. 
Finally, in the vertically hatched region $m_h < 100$~GeV which is excluded
by LEP2 searches for Higgs bosons~\footnote{A preliminary combined
limit of $\sim 106$~GeV (low
$\tan\beta$) and about 88~GeV (high $\tan\beta$) was presented at
SUSY2K~\cite{higgs} after our analysis was completed. This would enlarge
the vertically shaded region, but not affect the main results of the
paper which is the delineation of the region that may be explored at the
LHC. We expect qualitatively similar results say for $\tan\beta=5$.}.
We also show contours of $m_{\tg}=2$ TeV and $m_{\tu_R}=2$ 
TeV, to orient the reader as to the parameter space. Finally, the region to
the left of the dotted line is where two body decays of the neutralino
$\tz_2$ are allowed.

In Fig. \ref{amsb1}, parameter space points where there is a detectable
signal with 10 fb$^{-1}$ of integrated luminosity at the LHC are denoted
by triangles for the $0\ell$ channel, dots for the $1\ell$ channel,
circles for the $OS$ channel, squares for the $SS$ channel, and finally,
by stars for the $3\ell$ channel. Points that we explored, but found no
detectable signal in any of these channels, are denoted by asterisks.  In
the lower left, where $m_0<1000$ GeV and $m_{3/2}\le 60$ TeV, the SUSY
signal is visible in {\it all} channels considered. In this region, SUSY
particle production cross sections are high, and cascade decays lead to
visible signals in all channels. In the large $m_0\sim 2000$ GeV region
(where squarks are heavy) signals are observable in just the $0\ell$ and
$1\ell$ channels as anticipated above; the reach of the LHC with just 10
fb$^{-1}$ extends to $m_{3/2}\sim 60$ TeV, corresponding to values of
$m_{\tg}\sim 1350$ GeV.  
In the region of parameter space with small $m_0$ and large $m_{3/2}$,
the reach of the LHC extends up to $m_{3/2}\sim 110$ TeV, corresponding
to values of $m_{\tg}\sim 2200$ GeV. In this region, the maximum reach
is obtained in the $OS$ dilepton channel. The isolated leptons arise
from $\tg$ cascade decays to $\tst_1t$, and also from $\tq_R$ decays to
$\tz_2$ which have a branching fraction close to 100\%.  For parameters
to the left of the dotted line, $\tz_2$ always decays leptonically via
$\tz_2\to \ell\tell_R$ since sleptons are relatively light while squarks
are heavy.  ({\it e.g.}  $m_{\tell_R}\sim 300-400$ GeV, while
$m_{\tq_R}\sim 1500-2000$ GeV).  The $\tell_R\to \ell\tz_1$, which gives
rise to a hard isolated lepton.  In this region, the $\ell\bar{\ell}$
mass edge should be easily distinguished from background, and can serve
as the starting point for reconstructing sparticle masses in cascade
decays\cite{bcpt3l,lhc,ph,frank,lali,nojiri}.  

In Fig. \ref{amsb2}, we show the LHC reach again for $\tan\beta=3$, but
for $\mu >0$. For this sign of $\mu$ $m_h > 100$~GeV so there is no
vertically hatched region. For small values of $m_0 \alt 300$~GeV, where
the sneutrino is light, it is thus possible that the slanted hatched
region is slightly too big.
The reach results for the LHC are
qualitatively very similar to the $\mu <0$ case.

Finally, in Fig. \ref{amsb3}, we show the same parameter space plane but
for the large value of $\tan\beta =35$ and $\mu >0$. The results are
expected to be independent of the sign of $\mu$ when $\tan\beta$ is
large.  Again the slanted hatched region is what is excluded by the
chargino bound from LEP (where the same caveat as in Fig.~\ref{amsb2}
applies). The LEP limit on the $h$ mass does not provide any further
constraints. In the region of large $m_0$ and low $m_{3/2}$ REWSB is not
obtained. We see that the reach in $m_{3/2}$ for large $m_0$ is similar
to the low $\tan\beta$ cases, since sleptons and squarks are all very
heavy in this region. Multilepton signals are obtained for
$m_{3/2}=40$~TeV and very large $m_0$ from cascade decays of gluinos to
heavier charginos and neutralinos (which decay to real vector bosons)
plus third generation quarks: for low $\tan\beta$ these decays are
kinematically suppressed.
However, for low values of $m_0$, the reach in $m_{3/2}$
is considerably reduced compared to the low $\tan\beta$ case.
Here, values of $m_{3/2}\sim 90$ TeV are accessible with 10 fb$^{-1}$
of data, corresponding to $m_{\tg}\simeq m_{\tq}\sim 1850$ GeV.
The reduction is due mainly to the large $\tau$ Yukawa coupling, which
reduces the tau slepton mass compared to first and second generation
slepton masses. This enhances the decay of $\tz_2\to\ttau_1\tau$ at the expense
of $\tz_2$ decays to $e$s and $\mu$s. The $\tz_2$ originates 
again mainly via
cascade decay of $\tq_R$. The maximum reach for large $\tan\beta$ and
low $m_0$ is attained via $1\ell$, $OS$ and $3\ell$ channels.
In this case, the $3\ell$ events can come, for instance, from $\tg\tg$ 
events where $\tg\to\tst_1 t$, and 
where the $\tst_1$ has significant branching fractions into
$\tz_2$, and then into leptons.

The reach that we obtain (expressed in terms of the gluino and squark
masses) is similar to that in the mSUGRA framework. This is not
altogether surprising since the latter reach was mainly determined by
the $1\ell$ channel: while our arguments suggested that the multilepton
reach would be affected by the very different mass spectrum, there
appeared to be no reason why the reach in the $0\ell$ and $1\ell$
channels should qualitatively differ. By the same token, we might expect
that the overall reach would be relatively insensitive to variations of
the AMSB framework. The mass difference $m_{\tw_1}-m_{\tz_1}$ is,
however, {\it very sensitive} to our specific model assumptions: if
it is larger~\cite{frank} as, for instance, in the deflected AMSB
framework, then charginos would decay promptly, and its
decay products might be visible, making the signal more ``mSUGRA-like''.

In conclusion, the CERN LHC $pp$
collider should be able to discover supersymmetry with just 10 fb$^{-1}$ of 
integrated luminosity, over most of the mAMSB
parameter space where $m_{\tq}< 2$ TeV, or where $m_{\tg}<1300$ GeV
regardless of $m_{\tq}$.
Moreover, these signals should be observable via canonical
strategies~\cite{lhc,cms} advocated for SUSY searches in the mSUGRA
framework.
As in this case, over a large region of parameter space, there should be
observable signals in several channels.  Once a signal is found, then
signal events may be scrutinized for evidence of tracks from long lived
charginos, which occasionally penetrate into tracking chambers. These
tracks should either terminate, or result in a kink from the slow pion
daughter of the $\tw_1$.  Observability, if at all possible, will
require large event rates since $c\tau$ for the decaying chargino is
less than 5~$cm$ and we would be relying on charginos that live for
several lifetimes for this purpose: this feature cannot, therefore, be used to
extend the reach.  The rate will be sensitive to detector
characteristics and any instrumental backgrounds.  The terminating
tracks could serve to distinguish the mAMSB model from other scenarios.

%
\acknowledgments
This research was supported in part by the U.~S. Department of Energy
under contract number DE-FG02-97ER41022 and DE-FG-03-94ER40833.
%

%

\newpage
%
%

\iftightenlines\else\newpage\fi
\iftightenlines\global\firstfigfalse\fi
\def\dofig#1#2{\epsfxsize=#1\centerline{\epsfbox{#2}}}

%

%
\begin{figure}
\dofig{5in}{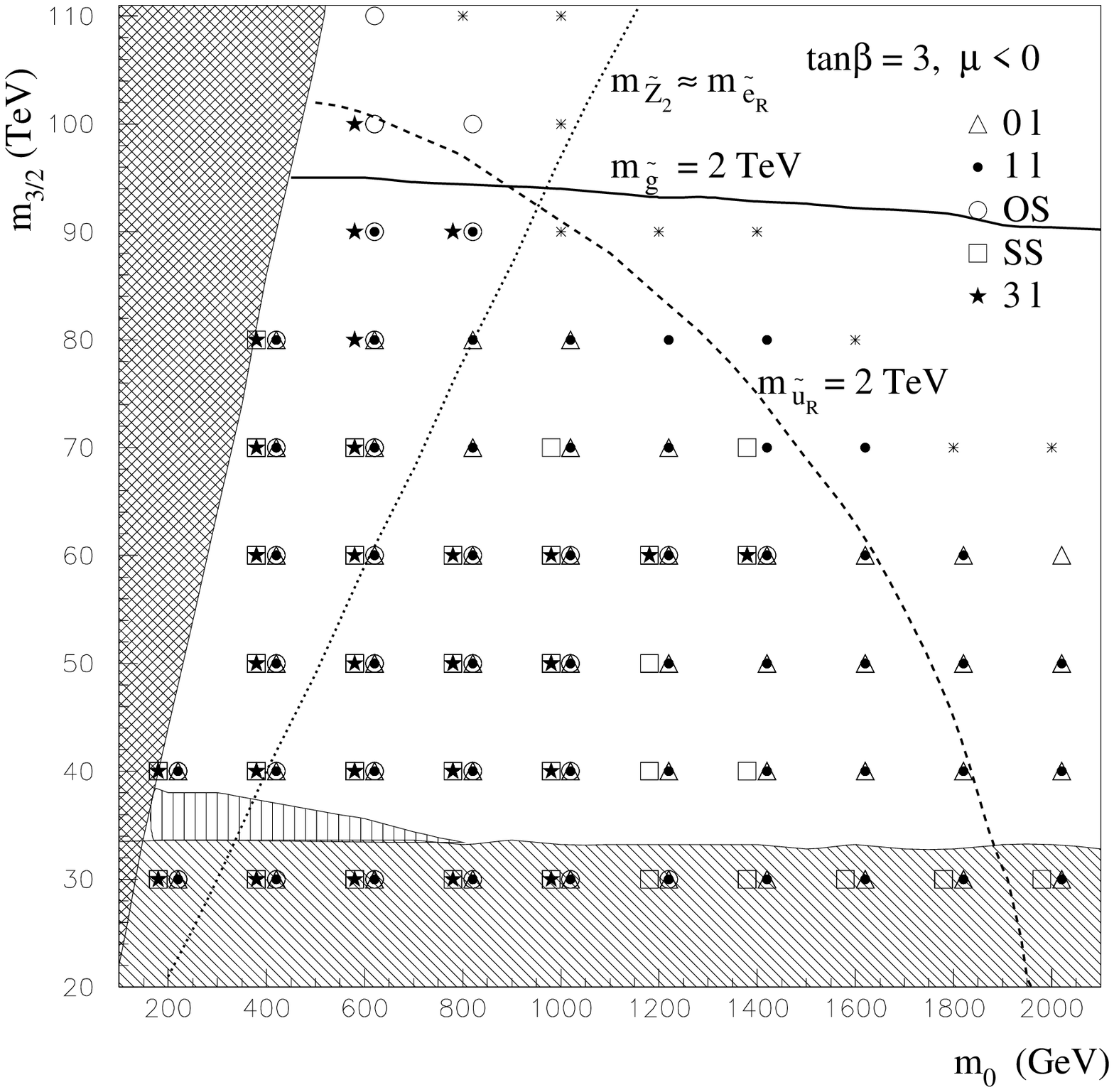}
\caption[]{A plot of the $m_0\ vs.\ m_{3/2}$ plane for $\tan\beta =3$
and $\mu <0$. The shaded regions are excluded by theoretical and
experimental constraints as discussed in the text. Each point in the
parameter space scanned is denoted by one or more symbols corresponding
to whether the SUSY signal for the point is  detectable
in any of the $0\ell$, $1\ell$, $OS$,
$SS$ or $3\ell$ channels at the CERN LHC, assuming
10 fb$^{-1}$ of integrated luminosity. Our criteria for detectability
are detailed in the text.  }
\label{amsb1}
\end{figure}
\begin{figure}
\dofig{5in}{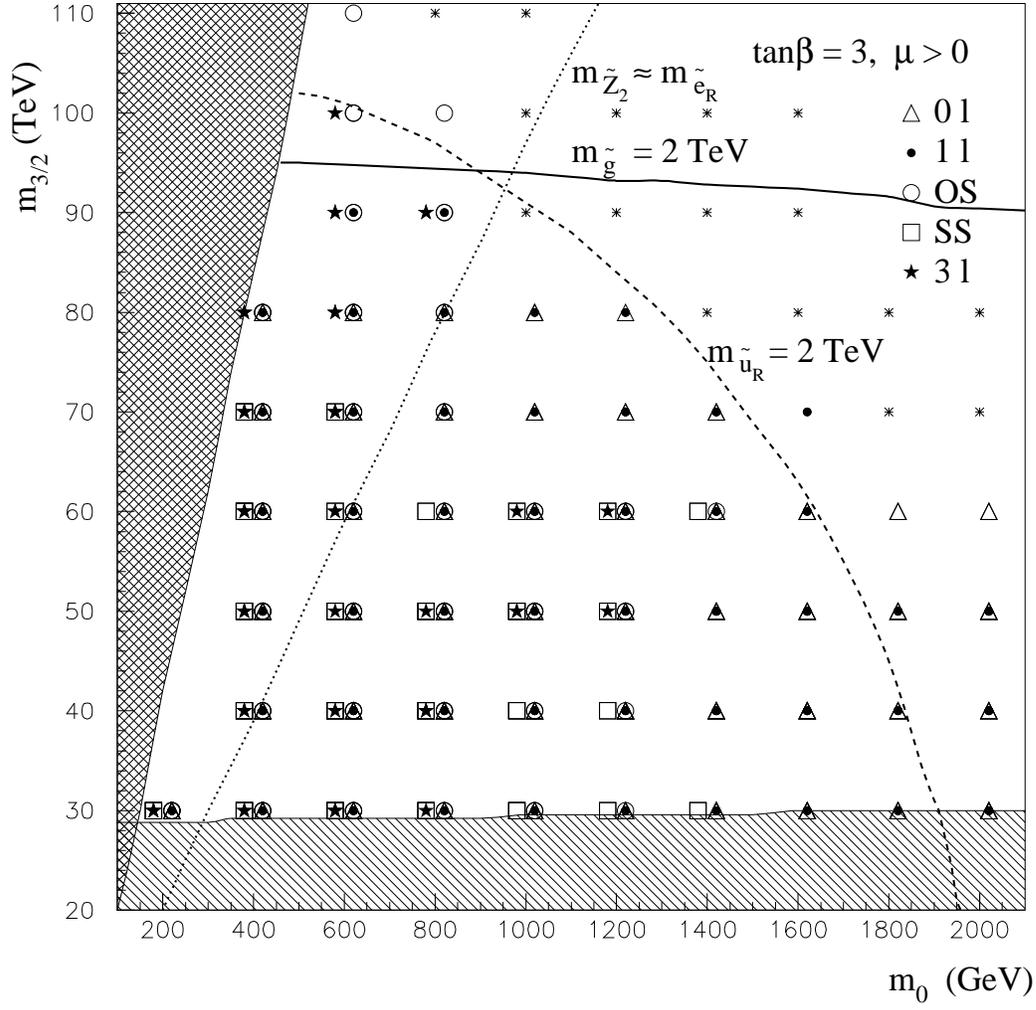}
\caption[]{Same as Fig. \ref{amsb1}, except we take $\mu >0$.}
\label{amsb2}
\end{figure}
\begin{figure}
\dofig{5in}{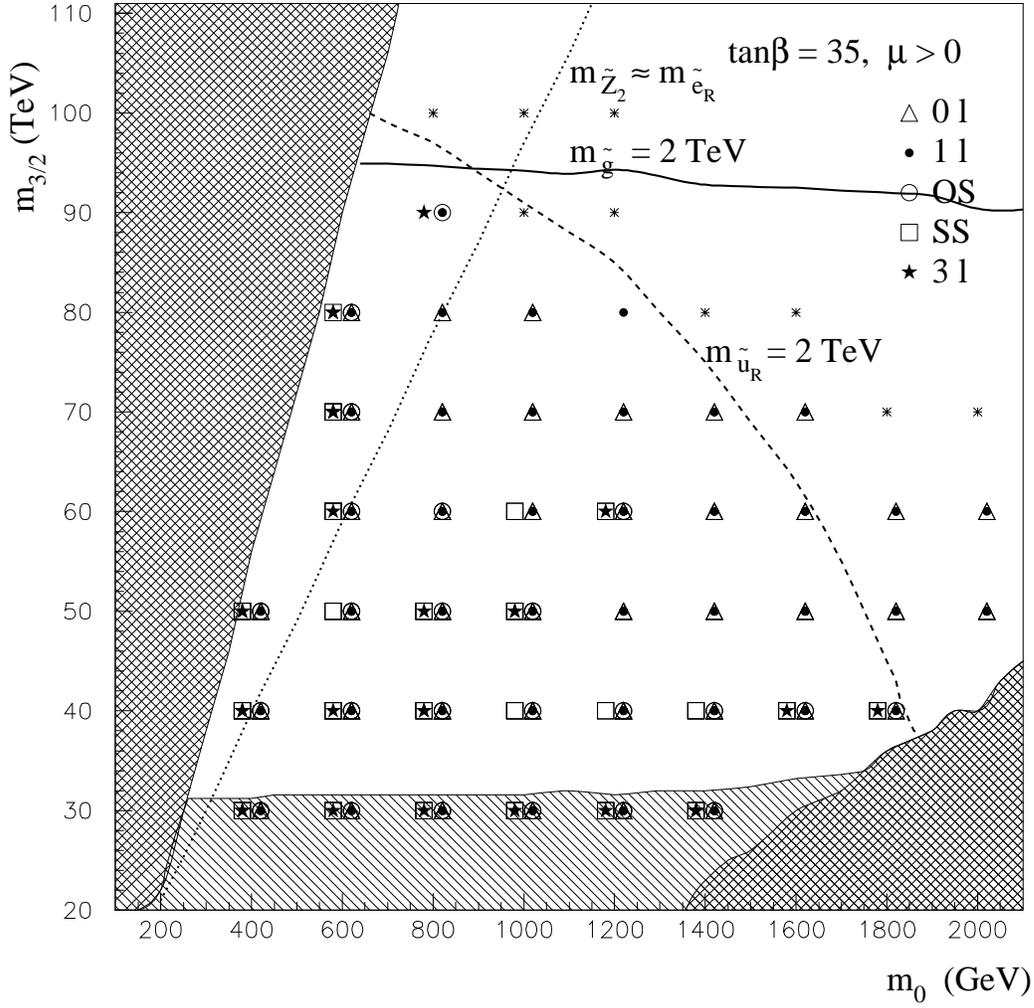}
\caption[]{Same as Fig. \ref{amsb1}, except we take $\tan\beta =35$, and
$\mu >0$.}
\label{amsb3}
\end{figure}

\end{document}